# Revealing of the new freedom of size: atomic energy levels, screening and charge transporting

Chang Q. Sun

E-mail: ecqsun@ntu.edu.sg; Fax: 65 6792 0415; http://www.ntu.edu.sg/home/ecqsun/

*School of Electrical and Electronic Engineering, Nanyang Technological University, Singapore 639798*

**Miniaturization of a solid forms a new freedom that is fascinating, which allows us not only to tune the physical properties of a solid but also enables us to gain information about the energy levels of an isolated atom and the effect of screening and charge transporting in reaction on the energy level shift, a calibration of crystal binding.**

Incorporating the recent bond order-length-strength correlation mechanism [Sun, *Phys. Rev. B* 69, 045105 (2004)] to the size dependence of Auger photoelectron coincidence spectra of Cu nanoparticles with and without being passivated has enabled us to gain quantitative information about the 2p and 3d level energies of an isolated Cu atom and their shift upon bulk formation. The developed approach also enabled us to discriminate the effect of crystal-field screening from the effect of valence recharging (due to surface passivation and substrate-particle interaction) on the binding energies to the electrons at different energy levels of a specimen.

## 1. Introduction

Crystal binding energy is a key quantity that determines the performance of a solid in aspects such as mechanical strength, thermal stability, lattice vibration, photon emission and absorption, electronic structures, magnetic, and dielectric performance. However, quantifying the crystal binding energy to electrons at different energy levels in a solid has been a long standing pursue. Determining the individual energy levels (intra-atomic trapping energy) of an isolated atom and their bulk shifts (inter-atomic binding) as well as the effects of crystal-field screening and valence recharging in reaction is also a high challenge.[1] With a combination of the most advanced laser cooling technology and X-ray photoelectron spectroscopy (XPS), one can measure the separation between two energy levels of the slowly moving gaseous atoms trapped by the laser beams.[2] What one can measure using an XPS are broad peaks of the core-levels resulting from contributions of atomic trapping, crystal binding and the effects of crystal orientation, surface relaxation, nanosolid formation or surface passivation. When a solid grows from atomic scale to macroscopic size with dimension $K_j$, the energy levels of an isolated atom expand to broad bands with prominent peaks that are conventionally referred as the core band characteristic energies. Where $K_j$ being the dimensionless form of size is the number of atoms lined along the radius of a spherical dot or cross the thickness of a plate. When the solid grows to form a prime unit cell or a monatomic chain,[3] the energy levels drop down abruptly to a minimum and then restore in a $K_j^{-1}$ fashion to the bulk value of the energy shift,[1] as illustrated in Figure 1. Such a core-level energy shift has been confirmed, for Cu example, using XPS or the Auger photoelectron coincidence spectroscopy (APECS) by a number of researchers.[4,5,6,7,8,9,10,11,12]

The observed energy shifts of the involved energy levels are often described with two concepts. One is the Auger parameter, $\alpha' = E_K + E_L,$[12] and the other is the Wagner plot, $\Delta E_K$ versus $\Delta E_L$.[13] The Auger parameter is the sum of the kinetic energy ($E_K$) of the Auger electrons and the binding (photoelectron) energy of electrons in the lower energy level ($E_L$). For Cu example: $\alpha' = E_k(LMM) + E_L(2p)$, which denotes the electronic polarizability of a chemical specimen.[12] As conventional practice, LMM represents the process of electron transition between the Cu-2p(L) and the degenerated Cu-3d(MM) levels. The energy shift of the Auger parameter is considered more sensitive to chemical reaction as it sums over the absolute shift of the $E_K$ and the $E_L$. The Wagner plot describes the correlation between the Auger kinetic energy shift $\Delta E_K$ and the photoelectron energy shift $\Delta E_L$ of specimens with different concentration (x) of the specific element such as Cu. The $\Delta\alpha'$ and the Wagner plot are given as:[12]

$$\begin{cases} \Delta\alpha' \cong 2\Delta R^{ex} & (Auger-parameter) \\ E_{K,x} - E_{K,Cu}(\infty) = \beta\left[E_{L,x} - E_{L,Cu}(\infty)\right] & (Wagner-plot) \end{cases}$$

(1)

The $\Delta\alpha'$ was assumed contribution equally from the involved two energy levels. $R^{ex}$ is the extra-atomic relaxation or polarization energy coming from the neighbors of the core-ionized atom. The coefficient $\beta$ is a constant and its value depends on the physical and chemical processes in which charge transportation and electron polarization occurs upon reaction.[14]

Conventionally, it is hardly possible to determine the binding energies of the two energy levels of an isolated atom simultaneously from a single APECS line. However, with two Auger parameters, for example, $\alpha'_1 = E_b(1s) + E_k(KLL)$ and $\alpha'_2 = E_b(2p) + E_k(KLL)$, measured from a number of compounds with different concentrations of the same element, one could estimate the separation between the two levels, $E_b(1s) - E_b(2p)$, but not the bulks shift of these two levels, in addition to the accuracy that is under question.[12] KLL is the Auger process between the K(s orbital) level and the generated L(p orbital) level. An approach is highly desirable for the absolute values of the APECS involved energy levels of an isolated atom and their bulk shift.

Wu, Garfunkel and Madey[8] deposited ultra thin Cu films on to $Al_2O_3$ substrate at 80 K and 300 K and then measured immediately their APECS data and found that the energy shifts of the 2p and 3d levels vary not only with the film thickness but also with substrate temperature. Yang and Sacher[11] reported a systematic study on the XPS $2p_{3/2}$-level and Auger $L_3M_{45}M_{45}$ line shift of Cu nanoparticles prepared under various deposition conditions on the substrates of highly-oriented pyrolitic graphite (HOPG) and Dow Cyclotene 3022 (CYL, a polymer of microelectronic industry interest), respectively. They demonstrated that the linear relations of size dependence of both the Cu-$2p_{3/2}$-level shift and the Auger parameter shift both of which follow the $K_j^{-1}$ scaling relation with a slope $B_\nu$ that is sensitive to processing conditions and substrate materials:

$$\Delta E_\nu\left(K_j\right) = \Delta E_\nu(\infty) + \frac{B_\nu}{K_j}$$

(2)

The subscript $\nu$ discriminates different energy levels, or the Auger parameter. These experimental findings show consistently that the Cu-2p and Cu-3d level shifts are both physical size and chemical treatment sensitive. Therefore, the size dependence of APECS shift should provide an opportunity for us to determine simultaneously the energies of the APECS involved levels and their shift and hence the crystal binding energy. Recently, by incorporating the bond order-length-strength (BOLS) correlation mechanism [1] and the size dependence XPS data, we determined the 2p level energy (-931.0 eV) of an isolated Cu atom and its bulk shift (-1.70 eV).[15] Approach has been also applied to the cases of Pd, Rh, Au, and Si atoms and their shift upon bulk formation and a number XPS data due to surface relaxation.[1] The objective of this presentation is to show that a further extension of the previous practice to the measured size dependence of the APECS lines allows us to gain information not only about the APECS involved energy levels of an isolated Cu atom but also the effects of crystal-field screening and valance recharging.

## 2. II  Principles

### 2.1 *Auger process: energy conservation*

Figure 1 illustrates the correlation between the APECS involved energy levels of Cu nanoparticles. In an Auger process, a photoelectron is emitted first from the lower $L(2p_{3/2})$ initial level upon the sample being excited, which leaves a hole behind the L level. This process of core-ionization causes a relaxation of all the energy levels due to the screen weakening of both the atomic potential to electrons in the outer shells and the crystal binding to the ion core. An electron transits then from the relaxed $M(3d_{5/2})$ upper final level to the hole in the lower L level. The energy released by this transition will kick an electron off the further relaxed M level to form the Auger electron. The Auger electron will overcome the binding energy $E_M$ and then escape from the solid with kinetic Auger energy $E_K$. If considering the absolute value of energies, the sum of $E_K$ and $E_M$ should equal to the energy difference between the L and the M levels. The vertical lines in Figure 1 denote their separation that follows energy conservation in the Auger process. Unlike the energy separation that is directly detectable, the absolute values of the $E_L(1)$ and the $E_M(1)$ and their bulk shifts $\Delta E_M(\infty)$ and $\Delta E_L(\infty)$ could not conventionally be measurable.

The energy conservation in the Auger process and the Auger parameter follow the relations (Figure 1):

$$\begin{cases} E_K\left(K_j\right) + E_M\left(K_j\right) = E_L\left(K_j\right) - E_M\left(K_j\right) & (Energy - conservation) \\ \alpha'\left(K_j\right) = E_K\left(K_j\right) + E_L\left(K_j\right) & (Auger - parameter) \end{cases} \tag{3}$$

From Figure 1, one may note that the energy shift of the upper M level is faster than the lower L level due to the different extent of crystal binding to the corresponding levels. According to the energy conservation and measurements,[11] the $\Delta E_K(K_j) < 0$ when the $E_L$ is enhanced.  In conventional practice, the $\Delta\alpha'(K_j)$ sums the absolute values of the $\Delta E_K(K_j)$ and the $\Delta E_L(K_j)$ and the reference point is chosen as the bulk shift $\Delta\alpha'(\infty) = 0$.[11] In real situation, the bulk shifts are non-zero, which is important to elucidate information behind the observations. Unfortunately, these bulk values could not be determined without using the freedom of changing size.[1]

Considering the absolute magnitudes of the energies and the referencing point to the specific energy level of an isolated atom, the magnitude of Auger parameter shift is actually twice the shift of the double-hole M level, $\Delta E_M$:

$$\begin{cases} \Delta E_K\left(K_j\right)+\Delta E_M\left(K_j\right)=\Delta E_L\left(K_j\right)-\Delta E_M\left(K_j\right) \\ \Delta\alpha'\left(K_j\right)=\left|\Delta E_K\left(K_j\right)\right|+\Delta E_L\left(K_j\right)=-\Delta E_K\left(K_j\right)+\Delta E_L\left(K_j\right) \end{cases}$$

$$\Delta\alpha'\left(K_j\right)=\Delta E_L\left(K_j\right)-\Delta E_K\left(K_j\right)=2\Delta E_M\left(K_j\right)$$

**(4)**

It is not surprising that the energy shifts of the APECS involved levels are not equal because of the effect of crystal-field screening that varies from level to level. For instance, electrons in the inner shells experience weaker crystal field compared with electrons in the outer shells due to the screening.

There are two possible ways to modulate the crystal binding energy. One is the bond-strength gain due to atomic coordination number (CN)-imperfection induced bond-length contraction and the other is bond nature alteration due to chemical passivation. On the other hand, chemical reaction, such as surface passivation and substrate-particle interaction, will transfer electrons from one specimen to another, which causes the valence charge redistribution.

### 2.2  BOLS correlation: Hamiltonian perturbation

According to the energy band theory, solid size reduction and surface passivation could perturb the overall potential in the Hamiltonian of an extended solid, which gives rise to the shift of an energy level denoted with arbitrary quantum number, $\nu$:[1,16]

$$\begin{cases} \Delta E_\nu\left(K_j\right)=\left[E_\nu(\infty)-E_\nu(1)\right]\left(1+\Delta_j\right) \\ \Delta_j=\sum_{i\leq 3}\gamma_{ij}\left(c_i^{-m}-1\right) \end{cases}$$

where

$$\begin{cases} \gamma_{ij}=\begin{cases} \tau c_i/K_j & \left(K_j>3\right) \\ 1 & \left(else\right) \end{cases} & \left(surface-to-volume\right) \\ c_i=2/\left\{1+\exp\left[\left(12-z_i\right)/\left(8z_i\right)\right]\right\} & \left(bond-contraction\right) \end{cases}$$

**(5)**

$E_\nu(1)$ is the core-level position of an isolated atom determined by the intra-atomic trapping potential. $\Delta E_\nu(\infty) = E_\nu(\infty) - E_\nu(1)$ is the bulk shift due to the crystal binding contribution. $\Delta_j$ originates from the CN-imperfection-induced bond contraction ($d_i = c_i d_0$) and the associated bond energy ($E_i = c_i^{-m} E_b$) enhancement in the outermost a few ($i \leq 3$) atomic layers. $E_b$ is the cohesive energy per coordinate. $\gamma_{ij}$ is the surface-to-volume ratio; the $c_i$ is the coordination ($z_i$) dependence of bond contraction coefficient. The BOLS coefficient, $m$, represents the nature of the bond. $\tau = 1$, 2, and 3 corresponds to the dimensionality of a thin plate, a rod and a spherical dot, respectively. Decoding the size dependence of measurement with the BOLS correlation has led to quantities of the vibration frequency of a Si-Si dimer,[17] the identities of a single Au-Au bond in gold monatomic chain,[3] and a C-C bond in carbon nantubes,[18] in addition to a single energy level of an isolated atom of a number of specimens.[1]

### 2.3 Extended Wagner plot

Although the absolute shift varies from level to level, the *relative* shift of each level to the bulk value of all the possible levels should be identical if no chemical process is involved. In order to discriminate the effect of crystal-field screening from the effect of valence recharging, we may extend the Wagner plot to correlate the relative shifts of the APECS involved two energy levels:

$$
\begin{cases}
\dfrac{E_M(K_j) - E_M(\infty)}{E_M(\infty) - E_M(1)} = \kappa_{ML} \dfrac{E_L(K_j) - E_L(\infty)}{E_L(\infty) - E_L(1)} & (Extended - Wagner - plot) \\
\eta_{ML} = \dfrac{E_M(\infty) - E_M(1)}{E_L(\infty) - E_L(1)} & (Screening - coefficient)
\end{cases}
\tag{6}
$$

The coefficient $\kappa_{ML}$ represents the effect of valence recharging due to chemical reaction on the shifts of the two energy levels. Ideally, $\kappa_{ML} = 1$, if no chemical reaction is involved. The screening coefficient, $\eta_{ML}$, represents the ratio of binding intensity to the two levels in a bulk of elemental solid. Therefore, the extended Wagner plot (Eq. (6)) would be more convenient and useful than the traditional one that correlates the absolute shifts of the Auger kinetic energy and the low-level binding energy, as the extended form could distinguish the effect of screen weakening from the effect of valence recharging on the relative shifts of the two levels for both a neutral and a chemically treated system.

## 3.　III　Results and discussion

It is possible now to decode the measured size dependent XPS $E_L(K_j)$ line, Auger parameter $\alpha'(K_j)$ with the above relations. Equilibrating the theoretical scaling relation (eq.(5)) to the measured size dependent APECS lines (Eq. (2)), we have,

$$
\Delta E_\nu(K_j) - \Delta E_\nu(\infty) = \begin{cases} \Delta E_\nu(\infty)\Delta_j \\ \dfrac{B_\nu}{K_j} \end{cases} \quad .
\tag{7}
$$

This relation leads to $\Delta E_\nu(\infty) \times \Delta_j = B_\nu / K_j$, or, $\Delta E_\nu(\infty) = E_\nu(\infty) - E_\nu(1) \equiv B_\nu(\Delta_j K_j)^{-1}$; where $E_\nu(\infty)$ and $E_\nu(1)$ are intrinsic values for the bulk and an isolated atom. $\Delta_j$ is bond nature, size and shape (m, $K_j$, $\tau$) dependent.[1] The $B_\nu$ value derived from measurement allows us to resolve the bulk shift, which is beyond currently available approaches.

The least-root-mean-square linearization of an APECS line from Cu on HOPG gives a slope $B_\nu$(= 2.0825) and the intercept (-0.0572) at $K_j^{-1} \to 0$. The intercept corresponds to the bulk value of $E_{2p}(\infty)$ = -932.7 eV.[19] By doing this, quantitative information about the atomic trapping ($2p_{3/2}$-level) energy, $E_L(1)$, of an isolated Cu atom, and the bulk bonding energy $\Delta E_L(\infty)$ to the $2p_{3/2}$ electrons have been obtained from the measurements, as given in Table 1. It is noted that all the levels will float up or down if the calibration (measured $E_L(\infty)$ value) changes, which vary slightly from source to source, but the bulk shifts are independent. For example, taking $E_L(\infty)$ = -932.1 eV,[20] will offset all the values by the same amount of 0.6 eV compared with that using $E_L(\infty)$ = -932.7 eV. Given the measured size dependent $E_L(K_j)$ and $\alpha'(K_j)$, quantities of $E_M(K_j)$ and $E_K(K_j)$ are readily available from the relation of energy conservation. Therefore, the $E_M(\infty)$ and $\Delta E_M(\infty)$, as well as the Auger kinetic energy $E_K(1)$ and $E_K(\infty)$ of the same Cu/HOPG sample can be obtained by fitting the corresponding size dependent APEC lines, as shown in Figure 2. Applying the same approach with the same bulk values of -1.70 and -2.12 eV for the $2p_{3/2}$ and $3d_{5/2}$ levels, respectively, to the APECS lines of other chemically treated samples has led to $m$ (Eq.(5)) values that vary with the process of chemical treatment, as indicated in Figure 3. Measurements show that the slope $B_\nu$ for the passivated Cu nanostructures varies from case to case due to the surface/interface bond strengthening (m value change) upon chemical passivation. The slope $\kappa_{ML}$ in the extended Wagner plots are independent of the shape and size of the particle but it is sensitive to the chemical treatment, demonstrating the effect of valence recharging due to passivation and substrate-particle interaction. The offsets of the straight lines in panel b and panel c of Figure 3 should be adjusted by replacing the bulk shifts of -2.12 and -1.70 eV with either known shifts of the corresponding compounds. Nevertheless, the compound bulk shifts should follow the relation:

$$\Delta E_M(\infty) / \Delta E_L(\infty) = (\eta\kappa)_{ML},$$

(8)

which is a joint contribution from the screening ($\eta_{ML} \cong 1.25$ is intrinsic for pure Cu bulk, which means the crystal binding to the M level is 25% stronger than the binding to the L level) and valence recharging.

Table 1 summarises information gained from the calculations. The determined $3d_{5/2}$ level,

-2.12 eV, is at the upper edge of the Cu 3d band. The Cu 3d band was determined to be 2.0 ~ 5.0 eV below the Fermi level using ultraviolet photoelectron spectroscopy (UPS) [21] and very-low-energy electron diffraction.[14] UPS spectra for Cu(110)-p(2 ×1)-O surface[22] shows the sharp peak at 2.1 eV below $E_F$, which may serve as direct evidence for the current determination of the bulk $3d_{5/2}$ level energy.

## IV Conclusion

In summary, we have described an approach of determining the two energy levels of an isolated Cu atom and their bulk shifts based on the size dependent APECS lines. Matching the BOLS predictions to the measured size dependence of the APECS lines gives the bulk shifts of the two levels of a metallic Cu and the bond strength of chemically treated surfaces. An extended Wagner plot has been developed, which correlates the relative energy shift of the upper final state to that of the lower initial state involved in APECS lines. This approach has enabled us to discriminate for the first time the effect of crystal-field screening from the effect of valence recharging on the binding energy of the electrons at the corresponding energy levels. Findings demonstrate further not only the impact of atomic CN imperfection and the new freedom of size but also the validity of the BOLS correlation mechanism.

Table and Figure captions

**Figure 1**

Schematic illustration of the size (with dimensionless form of size $K_j$) dependent energy-level shift and the relation of APECS $L_3M_{45}M_{45}$ lines: $(E_K + E_M = E_L - E_M)$. The energy levels, $E_M(K_j)$ and $E_L(K_j)$, and their separation, represented by the solid lines are measurable while the $E_M(1)$ and $E_L(1)$ and their bulk shift (dotted lines), can be obtained with the current approach, as summarized in Table 1. $E_L(2)$ corresponds to binding energy of two atoms.

**Figure 2**

Scaling relationship showing agreement between the predicted (solid line) and the measured size dependence of the $E_L$ and $E_M$ (scattered) and $E_K$ lines of Cu nanostructures deposited on HOPG substrate.[11] The vertical axis corresponds to $[E_v(K_j)-E_v(\infty)]/[E_v(\infty)-E_v(1)]$, $v$ represents L and M. Agreement gives the corresponding bulk shifts and the $\eta_{ML} = 1.25$ and $\kappa_{ML} = 1.05$, as given in Table 1.

**Figure 3**

Extended Wagner plots for Cu nanostructures on (a) HOPG and (b) CYL substrates with and without nitrogen (N) and Ar plasma sputtering upon deposition,[11] and on (c) $Al_2O_3$ substrate at 80 and 300 K.[8] $\kappa_{ML}$ is the slope of the correlation, which represents the valence recharging effect due to the plasma passivation and interfacial reaction. The index m is the parameter for the best fit of the $E_L(K_j)$ and $E_M(K_j)$ lines. Replacing the bulk shifts of -2.12 and -1.70 eV with those of the corresponding compounds (in panel b and c) could reshape the lines passing through the origin with a unity slope.

**Table 1**

The $3d_{5/2}$ and the $2p_{3/2}$ energies of an isolated Cu atom and their bulk shifts in magnitude derived from fitting the size dependent APECS lines of Cu nanostructures deposited on HOPG substrate (see **Figure 2**).   The monatomic $3d_{5/2}$ level is just below the work function 4.7~5.0 eV[21] and the bulk 3d band is within the range of 2 ~ 5 eV below the Fermi level. The screening coefficient $\eta_{ML}$ shows the relative intensity of the crystal binding to electrons in the two levels and the recharging coefficient κ is
sensitive to chemical process. Unit is in eV.

| $E_K + E_L$ | $\alpha'(\infty)$ | 1850.80[8,10,11] |
|---|---|---|
| $E_L(2p_{3/2})$ | $E_L(\infty)$ | -932.70[19] |
| | $E_L(1)$ | -931.00 |
| | $\Delta E_L(\infty)$ | -1.7 |
| $E_M(3d_{5/2})$ | $E_M(\infty)$ | -7.23 |
| | $E_M(1)$ | -5.11 |
| | $\Delta E_M(\infty)$ | -2.12 |
| $E_K(L_3M_{45}M_{45})$ | $E_K(\infty)$ | 918.24 |
| | $E_K(1)$ | 920.78 |
| | $\Delta E_K(\infty)$ | -2.54 |
| $\eta_{ML}$ | Screening coefficient | 1.25 |
| $\kappa_{ML}$ | Recharging coefficient | 1.05 |

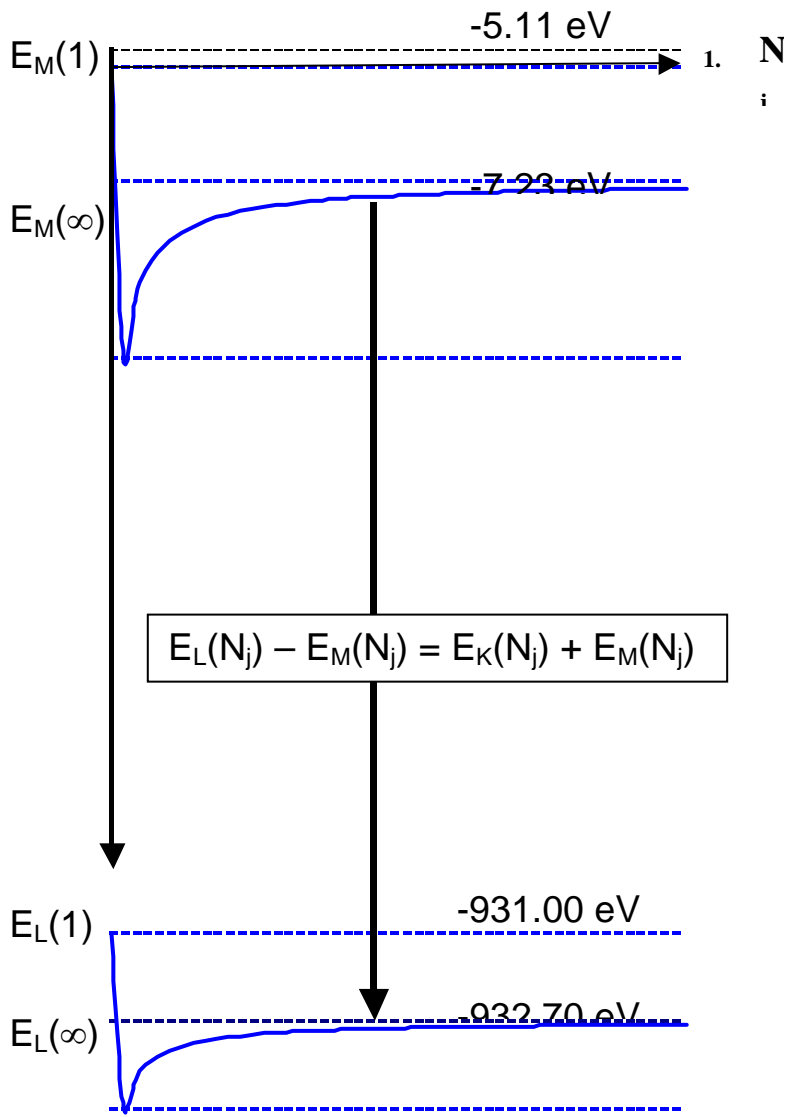

Fig. 1/CQ Sun

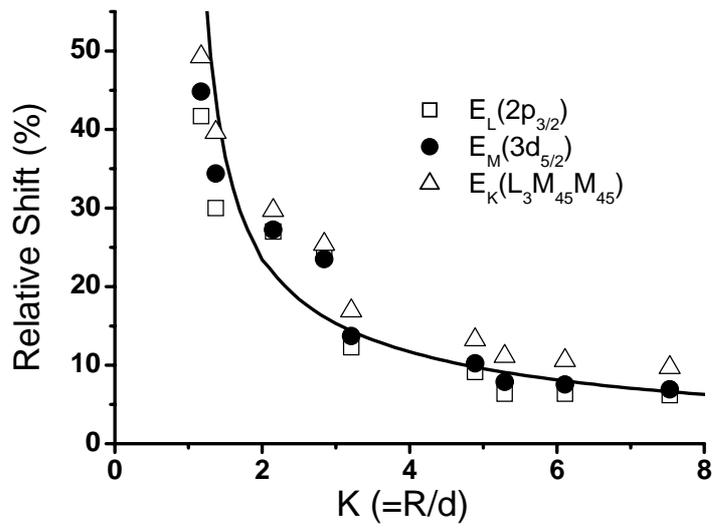

Fig. 2/CQ Sun

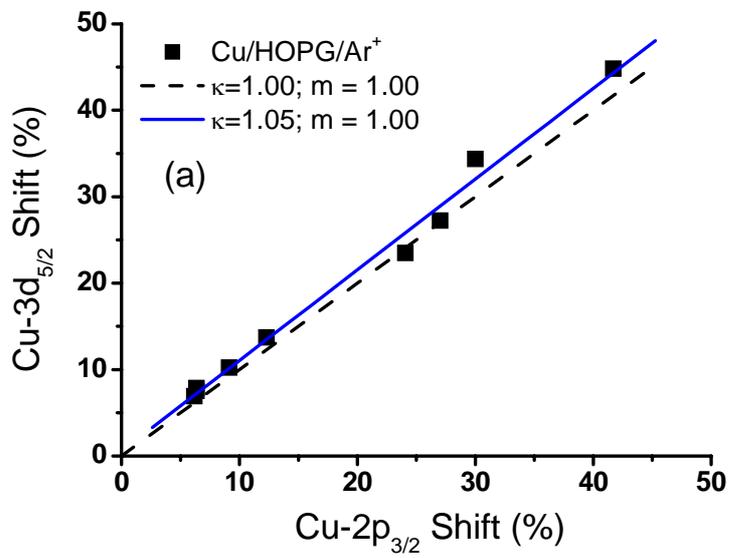

Fig. 3a/CQ Sun

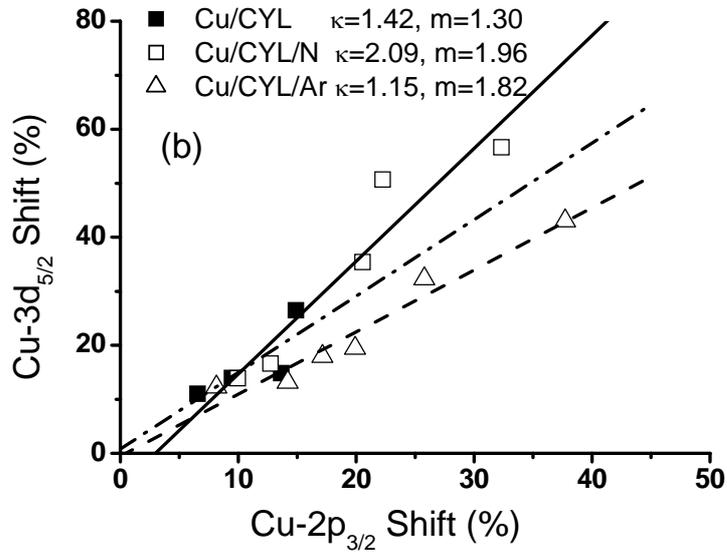

Fig. 3b/CQ Sun

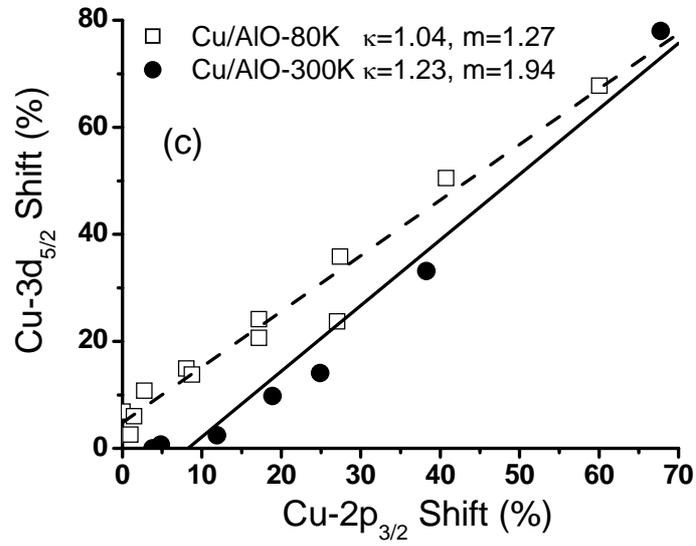

Fig. 3c/CQ Sun